\documentclass[runningheads]{llncs}

\usepackage[pdftex]{graphicx}
	\pdfoutput=1

\usepackage{amsmath,amssymb}

\usepackage{color}
	\definecolor{color1}{rgb}{0,0,0.8}

\usepackage[ruled]{algorithm2e}
  \SetKw{Break}{break}
  \SetFuncSty{textrm}
  \SetArgSty{textrm}
  \SetCommentSty{textit}
  \SetKwBlock{Forever}{repeat}{end}
  \SetKw{KwSide}{Side effects: }

\usepackage[colorlinks=true,
	linkcolor=black,
	citecolor=color1,
	filecolor=color1,
	menucolor=color1,
	urlcolor=color1,
	pdftex,
	pdfpagelayout=TwoColumnRight,
	pdftitle={Runtime-Flexible Multi-dimensional Arrays and Views for C++98 and C++0x},
	pdfsubject={Multi-dimensional Arrays},
	pdfauthor={Bjoern Andres},
	pdfcreator={Bjoern Andres},
	pdfproducer={Bjoern Andres},
	pdfkeywords={Array,Multi-dimensional,C++,C++0x,C++98}]{hyperref}

\usepackage[square,numbers,sectionbib]{natbib}

\begin{document}
\mainmatter
\title{Runtime-Flexible Multi-dimensional Arrays and Views for C++98 and C++0x}
\titlerunning{Multi-dimensional Arrays and Views for C++}
\authorrunning{B.~Andres et al.}
\author{Bjoern Andres, Ullrich Koethe, Thorben Kroeger, and Fred A. Hamprecht}
\institute{IWR, University of Heidelberg\\
\href{mailto:bjoern.andres@iwr.uni-heidelberg.de}{bjoern.andres@iwr.uni-heidelberg.de}} 

\maketitle

\begin{abstract}%
Multi-dimensional arrays are among the most fundamental and most useful data
structures of all. In C++, excellent template libraries exist for arrays whose
dimension is fixed at runtime. Arrays whose dimension can change at runtime
have been implemented in C. However, a generic object-oriented C++
implementation of runtime-flexible arrays has so far been missing. In this
article, we discuss our new implementation called Marray, a package of class
templates that fills this gap. Marray is based on views as an underlying
concept. This concept brings some of the flexibility known from script
languages such as R and MATLAB\textsuperscript{\textregistered} to C++. Marray
is free both for commercial and non-commercial use and is publicly available
from \url{www.andres.sc/marray}.
\end{abstract}
\section{Introduction and Related Work}
A $d$-dimensional array is a data structure in which each data item can be
addressed by a $d$-tuple of non-negative integers called coordinates.
Addressing data by coordinates is useful in many practical applications. As an
example, consider a digital image of 1920x1080 pixels. In this image, each
pixel can either be identified by the memory address where the associated color
is stored or, more intuitively, by a pair of coordinates
$(y, x) \in \{0, \ldots, 1919\} \times \{0, \ldots, 1079\}$.
Closely related to multi-dimensional arrays are multi-dimensional views. While
arrays are the storage containers for multi-dimensional data, views are
interfaces that allow the programmer to access data as if it was stored in an
array. In the above example, views can be used to treat any sub-image as if it
was stored in a separate array. Scientific programming environments such as R
\cite{ligges-2008,rproject} and MATLAB\textsuperscript{\textregistered}
\cite{matlab} exploit the versatility of views.

Some of the best implementations of arrays whose dimension is fixed at runtime
are written in C++, among these are the Boost Multidimensional Array Library
\cite{garcia-2005}, Blitz++ \cite{veldhuizen-1998}, and MultiArray of the image 
processing library Vigra \cite{vigra}. All three packages implement a common 
interface for views and arrays. Boost in addition allows the programmer to 
treat arrays as a hierarchy of nested containers. 
In a hierarchy of nested containers, an $(n+1)$-dimensional array is a 
container for $n$-dimensional arrays that have the same size. In this
hierarchy, 1-dimensional arrays differ from all other arrays in that they are 
containers of array \emph{entries} that need not be arrays themselves. This 
distinction is realized in all three implementations by means of template 
specialization with respect to the dimension of an array, an approach that 
achieves great runtime performance and compatibility with the simple 
multi-dimensional arrays that are native to C. However, template 
specialization also means that the data type of an array depends on its 
dimension. Thus, the hierarchy of nested containers does not generalize well in
C++ to arrays whose dimension is known only at runtime. This article therefore
presents an implementation that is based exclusively on views. Little is lost
because the hierarchy of nested containers can still be implemented as a cascade 
of views.

Many practical applications do not require runtime-flexibility because the 
dimensions of all arrays are either known to the programmer or restricted to a 
small number of possibilities that can be dealt with explicitly. However, the 
range of applications where the dimension of arrays is not known a priori and 
can change at runtime, possibly depending on the user input, is significant. In
particular, these are applications that deal with multi-variate data and/or 
multi-variate functions of discrete variables, e.g.~probability mass functions. 
It is no surprise that the runtime-flexible arrays of R and 
MATLAB\textsuperscript{\textregistered} have proven useful in these settings.

Section~\ref{section:concept} summarizes the mathematics of runtime-flexible 
multi-dimensional views and arrays. It is a concise compilation of existing 
ideas from excellent research articles
\cite{garcia-2005,veldhuizen-1998} and text books, e.g.~\cite{brass-2008}.
Section~\ref{section:implementation} deals with the C++ implementation of the
mathamtical concepts and provides some examples that show how the classes can
be used in practice. Readers who prefer a practical introduction are encouraged
to read Section~\ref{section:implementation} first. 
Section~\ref{section:implementation0x} discusses already implemented extensions
based on the C++0x standard proposal \cite{stroustrup-2006}.
Section \ref{section:conclusion} concludes the article.
\section{The Mathematics of Views and Arrays}
\label{section:concept}
\subsection{Views}
Views provide an interface to access data as if it was stored in an
array. A few definitions are sufficient to describe the properties
(syntax), function (semantics), and transformation of views. These
definitions are dealt with in this section. As they are implemented
one-to-one in the Marray classes, this section also explains in 
detail how these classes work internally.

\begin{definition}[View]
A non-degenerate multi-dimensional view is a quadruple $(d, s, t, p_0)
\in \mathbb{N} \times \mathbb{N}^d \times \mathbb{N}^d \times
\mathbb{N}$ in which $d$ is called the \emph{dimension}, $s$ the
\emph{shape}, $t$ the \emph{strides}, and $p_0$ the \emph{offset} of
the view. A tuple $(0, \emptyset, \emptyset, p_0)$ is called a
\emph{degenerate/scalar/0-dimensional} view. 
\end{definition}

Views allow the programmer to address data by tuples of $d$ positive
integers called coordinates. These coordinates are taken from ranges of
values that are determined by the view's shape:

\begin{definition}[Coordinates]
Given a view $V = (d, s, t, p_0)$,
\begin{equation}
C_V := \begin{cases}
\{0, \ldots, s_0-1\} \times \ldots \times \{0, \ldots, s_{d-1}-1\} & \text{if}\ d \not= 0\\
\emptyset\ \text{otherwise}
\end{cases}
\end{equation}
is called the set of coordinate tuples of $V$.
\end{definition}

According to this definition, coordinates start from 0 as is common in
C++, and not from 1 as in many script languages. Which data item is
addressed by a coordinate tuple $(c_0, \ldots, c_{d-1}) \in C_V$ is
determined by the addressing function of the view. This function is
parameterized by the view's strides and offset:

\begin{definition}[Addressing Function]
\label{definition:addressing-function}
Given a view $V = (d, s, t, p_0)$ with $d \not= 0$,
the function $a_V: C_V \rightarrow \mathbb{N}_0$ with
\begin{equation}
\forall c \in C_V: \quad a_V(c) = p_0 + \sum_{j = 0}^{d-1}{t_j c_j}
\end{equation}
is called the \emph{addressing function} of $V$.
\end{definition}

Semantically, a coordinate tuple $c = (c_0, \ldots, c_{d-1})$ identifies the 
data item that is stored at the address $a_V(c)$ in memory. Here are some 
examples: Assume that the integers $1, \ldots, 6$ are stored consecutively in
 memory at the addresses $100, \ldots, 105$. The six views in 
Tab.~\ref{table:view-examples}
address this memory and are written down next to the table in matrix notation,
i.e.~as tables in which the entry at row $j$ and column $k$
corresponds to the integer addressed by the coordinate $(j, k)$:

\begin{table}[b]
\caption{Multi-dimensional views on the same data can differ in
dimension, shape, strides, and offset.}
\label{table:view-examples}
\begin{tabular}{ccccc}
\hline
View & Dim $d$ & Shape $s$ & Strides $t$ & Offset $p_0$\\
\hline
$V_1$ & 2 & $(3, 2)$ & $(1, 3)$ & 100\\
$V_2$ & 2 & $(3, 2)$ & $(2, 1)$ & 100\\
$V_3$ & 2 & $(2, 3)$ & $(1, 2)$ & 100\\
$V_4$ & 2 & $(2, 3)$ & $(3, 1)$ & 100\\
$V_5$ & 2 & $(2, 2)$ & $(3, 1)$ & 101\\
$V_6$ & 1 & $(3)$    & $(2)$    & 101\\
\hline
\end{tabular}
\begin{minipage}{4cm}
\begin{eqnarray*}
V_1: \left( \begin{array}{cc} 1 & 4\\2 & 5\\3 & 6 \end{array} \right) & \quad
V_2: \left( \begin{array}{cc} 1 & 2\\3 & 4\\5 & 6 \end{array} \right) & \quad
V_3: \left( \begin{array}{ccc} 1 & 3 & 5\\2 & 4 & 6 \end{array} \right)\\
V_4: \left( \begin{array}{ccc} 1 & 2 & 3\\4 & 5 & 6 \end{array} \right) & \quad
V_5: \left( \begin{array}{ccc} 2 & 3\\5 & 6 \end{array} \right) & \quad
V_6: (2, 4, 6)
\end{eqnarray*}\\

\end{minipage}
\end{table}

The views $V_1, \ldots, V_4$ address the same set of integers but in a
different shape and with different addressing functions. Perhaps more
interestingly, $V_5$ is a sub-view of $V_4$ that has the same
dimension but a different shape, and $V_6$ is a sub-view of $V_3$
whose dimension has been reduced. In general, sub-views can be defined
as follows:

\begin{definition}[Sub-View]
\label{definition:sub-views}
Given a view $V = (d, s, t, p_0)$ with $d \not= 0$, a start coordinate
$c \in C_V$, and a shape $s' \in \mathbb{N}^d$ such that
$\forall j \in \{0, \ldots, d-1\}: c_j + s_j' \leq s_j$,
\begin{equation}
\text{sub-view}(V, c, s') := (d, s', t, p_0 + a_V(c))
\end{equation}
is called the \emph{sub-view of $V$ with the shape $s'$, starting at
the coordinate $c$}.
\end{definition}

The convenient access to sub-views is one of the main reasons why
multi-dimensional views are useful in practice.

As important as the construction of sub-views is the binding of
coordinates. If one coordinate in a $d$-dimensional view is bound to a
value, the result is a $(d-1)$-dimensional view. In the above example,
$V_6$ arises from $V_3$ by binding coordinate 0 to the value 1. In
general, coordinate binding works as follows:

\begin{definition}[Coordinate Binding]
Given a view $V = (d, s, t, p_0)$ with $d \not= 0$, a dimension
$j \in \{0, \ldots, d-1\}$ and a value $x \in \{0, \ldots, s_j-1\}$,
\begin{equation}
\text{bind}(V, j, x) := (d-1, s', t', p_0')
\end{equation}
with $s' = (s_0, \ldots, s_{j-1}, s_{j+1}, \ldots, s_{d-1})$,
$t' = (t_0, \ldots, t_{j-1}, t_{j+1}, \ldots, t_{d-1})$ and
$p_0' = a_V(c)$ with $c \in C_V$ such that
$\forall k \in \{0, \ldots, d-1\}: c_k = x \delta_{j k}$ is said to
arise from $V$ by \emph{binding} coordinate $j$ to the value $x$.
\end{definition}

By Def.~\ref{definition:sub-views}, sub-view$(V, c, s')$ has the same dimension
as $V$. However, the shape of the sub-view may be equal to one in some 
dimensions, i.e.~$s_j' = 1$ for some $j$. Since 0 is the only admissible 
coordinate in these singleton dimensions, it makes sense to bind such 
coordinates to 0. Binding the coordinates in all singleton dimensions to 0 is 
called \emph{squeezing}.

An operation that preserves both the dimension and the memory
addressed by a view is permutation. Permuting a view permutes the
view's shape and strides, respectively:

\begin{definition}[Permutation]
The \emph{permutation} of a non-degenerate view $V = (d, s, t, p_0)$
w.r.t.~a bijection
$\sigma: \{0, \ldots, d-1\} \rightarrow \{0, \ldots, d-1\}$
is the view
\begin{equation}
\text{permute}(V, \sigma) := (d, s', t', p_0)
\end{equation}
where $s', t' \in \mathbb{N}^d$ and
$\forall j \in \{0, \ldots, d-1\}: s'_j = s_{\sigma(j)} \wedge t'_j = t_{\sigma(j)}$.
\end{definition}

Two special cases of permutations are \emph{transpositions} and
\emph{cyclic shifts}. Transpositions exchange the shape and strides in
only two dimensions. In the above example, $V_1$ and $V_4$ are
transposes of each other, and so are $V_2$ and $V_3$. Cyclic shifts
permute a view in a cyclic fashion. As an example, consider a
3-dimensional view whose shape is $(2,3,7)$. If this view is shifted
by 1, the resulting view has the shape $(7, 2, 3)$, and a shift by
-1 yields a view having the shape $(3, 7, 2)$. In general, cyclic
shifts can be defined and computed as follows:

\begin{definition}[Cyclic Shift]
The cyclic shift of a non-degenerate view $V = (d, s, t, p_0)$ w.r.t.
$z \in \mathbb{Z}$ is the view
\begin{equation}
\text{shift}(V, z) :=
\begin{cases}
\text{shift}(V, z\ \text{mod}\ d) & \text{if}\ d \leq |z|\\
\text{shift}(V, z - d)            & \text{if}\ 0 < z < d\\
(d, s', t', p_0)                  & \text{otherwise}
\end{cases}
\end{equation}
with $s', t' \in \mathbb{N}^d$ and $\forall j \in \{0, \ldots, d-1\}:
s'_j = s_{(j-z)\ \text{mod}\ d} \wedge
t'_j = t_{(j-z)\ \text{mod}\ d}$.
\end{definition}
\subsection{Scalar Indexing and Iterators}
The coordinate tuples of a view can be put in some order. Imposing
such an order allows the programmer to access any data item under the view by a
single index, namely the index of the associated coordinate tuple in
the given order. This is useful in practice because it in turn allows
us to handle sub-views as if they were single-indexed containers
holding a subset of data. Moreover, it facilitates the definition of
iterators \cite{austern-1998} on views.

Among all possible orders that can be imposed on coordinate tuples, two
are most commonly used\footnote{Note, however, that more complex orders can be
obtained by defining views with specific strides.}. In the First Coordinate
Major Order (FCMO), the first coordinate is used as the strongest ordering
criterion, meaning that one tuple is greater than all tuples whose first
coordinate is smaller. Coordinates at higher dimensions are used for ordering
only if all coordinates at lower dimensions are equal. In the Last Coordinate
Major Order (LCMO), the last coordinate is the strongest ordering criterion.
In the special case of 2-dimensional views, FCMO and LCMO are called 
\emph{row-major order} and \emph{column-major order}, respectively. These terms
refer to the matrix notation of data under 2-dimensional views. FCMO is used in 
native C arrays whereas LCMO is used in Fortran and MATLAB. Both orders are 
defined implicitly by a function that maps coordinate tuples to unique integer 
indices. One coordinate is smaller than another precisely if the associated 
index is smaller.

\begin{definition}[Indexing]
\label{definition:indexing}
Given a view $V = (d, s, t, p_0)$ with $d \not= 0$ and a coordinate
$c = (c_0, \ldots, c_{d-1}) \in C_V$,
\begin{eqnarray}
\text{fcmo}(c) := \sum_{j = 0}^{d-1}{ u_j c_j }
& \text{with} & u_j = \prod_{k=j+1}^{d-1}{ s_k } \enspace ,\\
\text{lcmo}(c) := \sum_{j = 0}^{d-1}{ u_j c_j }
& \text{with} & u_j = \prod_{k=0}^{j-1}{ s_k } \enspace .
\end{eqnarray}
are called the FCMO- and LCMO-\emph{index} of $c$, respectively. Given
that either FCMO or LCMO is used, $u_0, \ldots, u_{d-1}$ are called
the \emph{shape strides} of $V$.
\end{definition}

As an example, consider a 3-dimensional view $V = (d, s, t, p_0)$.
Herein, the indices that correspond to a given coordinate $c \in C_V$
are computed according to
\begin{eqnarray*}
\text{fcmo}(c) & = & s_1 s_2 c_0 + s_2 c_1 + c_2 \enspace ,\\
\text{lcmo}(c) & = & c_0 + s_0 c_1 + s_0 s_1 c_2 \enspace .
\end{eqnarray*}

The index that corresponds to a coordinate tuple can be computed
according to Def. \ref{definition:indexing}. Conversely, the
coordinates that correspond to a given FCMO- or LCMO-index are
computed by means of Alg.~\ref{algorithm:index-to-coordinates}.
Given that either FCMO or LCMO is used, it can happen that the strides
are equal to the shape strides of a view. Such views are called
\emph{unstrided}. In an unstrided view $V = (d,s,t,p_0)$, the address
that corresponds to an index $x \in \mathbb{N}_0$ is simply $x + p_0$,
whereas in a strided view, one needs to compute first the coordinate
$c$ that corresponds to the index $x$ (Alg. \ref{algorithm:index-to-coordinates})
and then the address $a_V(c)$ (Def. \ref{definition:addressing-function}).

\begin{algorithm}[h]
\label{algorithm:index-to-coordinates}
\caption{IndexToCoordinates}
\KwIn{$x \in \mathbb{N}_0$ (index), $(u_0, \ldots, u_{d-1}) \in \mathbb{N}^d$ (shape strides)}
\KwOut{$(c_0, \ldots, c_{d-1}) \in \mathbb{N}^d$ (coordinates)}
\eIf{$u_0 = 1$}{
	// LCMO\\
	\For{j = d-1 \KwTo 0}{
		$c_j \leftarrow \lfloor x / u_j \rfloor$\;
		$x \leftarrow x\ \text{mod}\ u_j$\;
	}
}{
	// FCMO\\
	\For{j = 0 \KwTo d-1}{
		$c_j \leftarrow \lfloor x / u_j \rfloor$\;
		$x \leftarrow x\ \text{mod}\ u_j$\;
	}
}
\end{algorithm}

In summary, we have seen that views are powerful interfaces to address
data either by coordinates or by single indices. It is simple to
obtain sub-views and to bind and permute coordinates.
\subsection{Arrays}
\label{section:arrays}
A multi-dimensional array is a data structure whose interface is a
view. While views only reference data via their addressing function,
arrays contain data. In the following definition, the memory is 
modeled as a function $\mu$ that maps addresses to memory content.

\begin{definition}[Array]
A $d$-dimensional array is a tuple $(V, q, \mu)$ such that
$V = (d, s, t, p_0)$ is a view,
$q \in \{\textnormal{FCMO, LCMO}\}$,
$V$ is unstrided w.r.t.~$q$, and $\mu$ is a function
\begin{equation}
\mu: \left\{p_0,\ \ldots,\ p_0 + \left(\prod_{j = 0}^{d-1}{s_j}\right) - 1 \right\} \rightarrow \mathbb{N} \enspace .
\end{equation}
For each $c \in C_V$, $\mu(a_V(c))$ is called the \emph{entry} of the array at position $c$.
Moreover, $|C_V|$ is termed the array's \emph{size}.
\end{definition}

Two transformations are defined on arrays, namely reshaping and
resizing. Reshaping can change the dimension and shape of an array
while preserving its size and entries.

\begin{definition}[Reshaping]
Given an array $A = ((d, s, t, p_0), q, \mu)$ as well as
$d' \in \mathbb{N}$, and $s' = (s_0, \ldots, s_{d'-1})$ such that
$\prod_{j=0}^{d'-1}{s'_j} = \prod_{j=0}^{d-1}{s_j}$, the
\emph{reshaping} of $A$ w.r.t. $s'$ is the array
\begin{equation}
\text{reshape}(A, s') := ((d, s', t', p_0), q, \mu)
\end{equation}
in which $(d, s', t', p_0)$ is a view that is unstrided w.r.t.~$q$.
\end{definition}

In fact, reshaping can not only be defined for arrays but also, more
generally, for unstrided views. 

In constrast to reshaping, resizing can change the size and hence the 
interval of memory of an array:

\begin{definition}[Resizing]
Given an array $A = (V, q, \mu)$, a new dimension $d' \in \mathbb{N}$
and a new shape $s' = (s_0, \ldots, s_{d'-1})$, an array $(V', q, \mu')$ 
is called a resizing of $A$ w.r.t.~$s'$, denoted resize$(A, s')$, if and 
only if the following conditions hold:

(i) $V' = (d', s', t', p_0')$ is a view that is unstrided w.r.t. $q$.
(Note that the offset $p_0'$ of the new array can differ from that of 
$V$ due to a possible re-allocation of memory).

(ii) entries of $A$ are preserved according to the following rule:
\begin{eqnarray}
\forall (c, c') \in D: \quad \mu(a_V(c)) = \mu'(a_{V'}(c'))
\end{eqnarray}
with
\begin{eqnarray*}
D = \{
(c, c') \in C_V \times C_V'\ |\
\forall j \in \{0, \ldots, \text{min}(d,d') - 1\}: c_j = c_j' \\
\wedge \forall j \in \{\text{min}(d,d'), \ldots, d - 1\}: c_j = 0\\
\wedge \forall j \in \{\text{min}(d,d'), \ldots, d' - 1\}: c_j' = 0 )
\}
\end{eqnarray*}
\end{definition}

Finally, all transformations of views can be used similarly with
arrays.
\section{Implementation}
\label{section:implementation}
The definitions introduced above are implemented in C++ in the Marray package
\cite{marray}. Marray depends only on the C++ Standard Template Library (STL)
\cite{austern-1998}. The single header file \verb$marray.hxx$ is sufficient to
use the package. This header file contains the source code as well as reference
documentation in the doxygen format \cite{doxygen}. In addition to  this file,
we provide unit tests \cite{hamill-2004} in the file \verb$tests.cxx$ as well as
the reference documentation in HTML.

\begin{figure}
\centering
\includegraphics{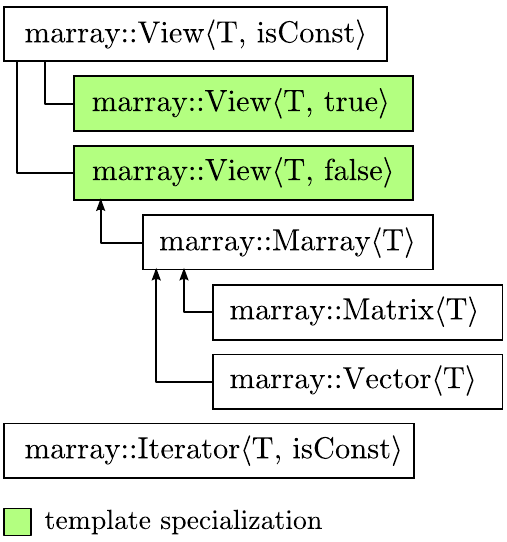}
\caption{Class template hierarchy of the Marray package. The five
major class templates are View, Marray, Matrix, Vector, and Iterator.
The Boolean template parameter \emph{isConst} is used to determine 
whether the data addressed by views and iterators is constant or 
mutable.}
\label{figure:classes}
\end{figure}

Five major class templates are defined in the namespace \verb$marray$.
These are \verb$View$, \verb$Marray$, \verb$Matrix$, \verb$Vector$,
and \verb$Iterator$. Their organization is depicted in Fig.~\ref{figure:classes}.
The Boolean template parameter \verb$isConst$ is
used to determine whether the data addressed by views and iterators
is constant or mutable. This facilitates a unified implementation for
both cases without any redundancy in the code, cf. \cite{meyers-2005}.
The class templates \verb$Marray$, \verb$Matrix$, and \verb$Vector$
inherit the interface from \verb$View<T, false>$.
\subsection{Using Arrays}
The simplest means to construct an array is a pair of iterators that
point to the beginning and the end of a sequence that determines the
array's shape:
\begin{verbatim}
size_t shape[] = {3, 2, 4};
marray::Marray<float> a(shape, shape+3);
\end{verbatim}
Constructing matrices and vectors is even simpler and works as most
programmers will expect, namely by providing the size and the number
of rows and columns, respectively:
\begin{verbatim}
marray::Vector<float> v(42);
marray::Matrix<float> m(7, 8);
\end{verbatim}
In addition to the shape, one can specify an initial value for all
array entries as well as the order in which entries are stored, e.g.
\begin{verbatim}
marray::Marray<float> b(shape, shape+3, 1.0f,
   marray::FirstMajorOrder);
\end{verbatim}
By default, all entries of a \verb$marray::Marray<T>$ are initialized
with \verb$T()$ and are stored in Last Coordinate Major Order, cf.
Section \ref{section:concept}. Depending on the application, the
initialization of array entries is sometimes unnecessary and can thus
be skipped to improve performance. Initialization skipping works as
follows:
\begin{verbatim}
size_t shape[] = {3, 2, 4};
marray::Marray<float> a(marray::SkipInitialization,
   shape, shape+3);
marray::Vector<float> v(marray::SkipInitialization, 42);
marray::Matrix<float> m(marray::SkipInitialization, 7, 8);
\end{verbatim}

After construction, the dimension, size, shape, and storage order of
an array can be obtained as follows.
\begin{verbatim}
unsigned short dimension = a.dimension();
size_t size = a.size();
bool firstMajorOrder = a.firstMajorOrder();
marray::Vector<size_t> shape(dimension);
for(size_t j=0; j<dimension; ++j)
   shape[j] = a.shape(j);
\end{verbatim}

The entries of an array can be accessed in three different ways: by
coordinates, by single indices, and by means of STL compliant random
access iterators, cf. \cite{austern-1998}. In fact, the following
four assignments have the same effect on the array \verb$a$.
\begin{verbatim}
// 1.
a(1, 0, 2) = 4.2f;
// 2.
size_t pos[] = {1, 0, 2};
a(pos) = 4.2f;
// 3.
a(13) = 4.2f;
// 4.
marray::Marray<float>::iterator it = a.begin();
it[13] = 4.2f;
\end{verbatim}

It can sometimes be useful to print the entries of an array
to \verb$std::cout$. This can be done using
\begin{verbatim}
std::cout << a.asString(marray::TableStyle);
std::cout << a.asString(marray::MatrixStyle);
std::cout << a.asString(); // MatrixStyle is the default
\end{verbatim}
In table style output, each printed row consists of a coordinate
tuple and the corresponding array entry. In the more compact matrix
notation only the entries of the array are printed.

Both the shape and the size of an array can be changed at runtime.
Reshaping modifies an array's shape and dimension while preserving
its size. Resizing can in addition cause the amount of memory
allocated by the array to grow or shrink.
\begin{verbatim}
size_t newShape[] = {2, 2, 3, 2};
a.reshape(newShape, newShape+4);
newShape[0] = 4;
a.resize(newShape, newShape+4);
\end{verbatim}

The function \verb$resize$ can alternatively be called with a third 
parameter that specifies the initial value for newly allocated entries.
For matrices and vectors, reshaping and resizing works as follows:
\begin{verbatim}
v.resize(56);
m.reshape(8, 7);
m.resize(2, 4);
\end{verbatim}

It can sometimes be useful to permute the dimensions of an array, e.g.~to 
transpose a matrix. Three functions, \verb$permute$, \verb$transpose$, and 
\verb$shift$ serve this purpose. While \verb$permute$ deals with the most 
general case of permuting dimensions in any desired way, \verb$transpose$ with
two parameters swaps any two dimensions, \verb$transpose$ with no parameters
reverses the order of dimensions, and \verb$shift$ shifts them in a cyclic 
fashion. No matter which function is used, only the array's interface is 
adjusted; no data is moved or copied.
\begin{verbatim}
size_t shape[] = {3, 2, 4};
marray::Marray<float> c(shape, shape+3);
size_t permutation[] = {1, 0, 2};
c.permute(permutation); // (2, 3, 4)
c.transpose(0, 2); // (4, 3, 2)
c.shift(-1); // (3, 2, 4)
c.shift(2); // (2, 4, 3)
c.transpose(); // (3, 4, 2)
\end{verbatim}

Finally, the arithmetic operators \verb$+$, \verb$-$, \verb$*$,
\verb$/$, \verb$+=$, \verb$-=$, \verb$*=$, \verb$/=$ are defined.
They operate on an array and its entry data type (in any order),
as well as on pairs of arrays that have the same shape. In the
latter case, the operation is performed on each pair of entries,
for every coordinate. In summary, this allows the programmer to
use arithmetic expressions like these:
\begin{verbatim}
marray::Marray<float> d;
d = -a + 0.5f*a - 0.25f*a*a;
d = 1.0f / (1.0f + a*a);
d = (a /= 2.0f);
--a;
\end{verbatim}
\subsection{Using Views}
Arrays, including matrices and vectors, are containers. Views are interfaces 
that allow the programmer to access data as if it was stored in an array. A 
view can be constructed either as a sub-view of another view or array, or 
directly on an interval of memory. In the following example, a 2-dimensional 
sub-view is constructed that ranges from position $(3, 2, 4)$ to position 
$(7, 2, 8)$ in a 3-dimensional array.
\begin{verbatim}
size_t shape[] = {20, 20, 20};
marray::Marray<float> d(shape, shape+3);
size_t base[] = {3, 2, 4};
size_t subShape[] = {5, 1, 5};
marray::View<float> v = d.view(base, subShape);
v.squeeze(); // collapse singleton dimension
\end{verbatim}

Each view defines an internal order of coordinates, either First or
Last Coordinate Major Order. This order determines how an iterator
traverses the view as well as how single indices are mapped to
coordinates, e.g.~which entry of \verb$d$ in the above example is
referenced by, say, \verb$v(7)$. The coordinate order of a sub-view
need not be the same as the coordinate order of the view based on
which it is constructed, although this is the default. Instead, it is
possible to specify the coordinate order of sub-views explicitly, e.g.
\begin{verbatim}
marray::View<float> v = 
   d.view(base, subShape, marray::FirstMajorOrder);
\end{verbatim}
This facilitates the construction of sub-views that behave exactly
like the views or arrays on which they are based, except that the
coordinate order is reverted:
\begin{verbatim}
marray::Vector<size_t> base(d.dimension());
marray::Vector<size_t> subShape(d.dimension());
for(size_t j=0; j<d.dimension(); ++j)
   subShape(j) = d.shape(j);
marray::View<float> v = d.view(base.begin(), subShape.begin(),
   marray::FirstMajorOrder);
\end{verbatim}

Views can be constructed directly on an interval of memory. If all data
in this interval is to be referenced by the view, i.e.~if the view is
to be unstrided (cf. Section \ref{section:concept}), it is sufficient to
provide the view's shape and a pointer to the beginning of the data.
\begin{verbatim}
float data[24];
size_t shape[] = {3, 2, 4};
marray::View<float> w(shape, shape+3, data);
\end{verbatim}
The same constructor can be used with two additional parameters,
\begin{verbatim}
marray::View<float> w(shape, shape+3, data,
   marray::LastMajorOrder, marray::FirstMajorOrder);
\end{verbatim}

These parameters specify the external coordinate order based on which
the strides of the view are computed as well as the internal
coordinate order that is used for indexing and iterators. By default,
Last Coordinate Major Order is used for both. Views on constant data
are constructed similar to views on mutable data, e.g.
\begin{verbatim}
marray::View<float, marray::Const> w(shape, shape+3, data);
\end{verbatim}
Constructing unstrided views is only the simplest case. In general,
the strides as well as the offset of a view (cf.~Section
\ref{section:concept}) can be set explicitly, e.g.
\begin{verbatim}
size_t shape[] = {3, 2, 4};
size_t strides[] = {2, 1, 6};
size_t offset = 0;
marray::View<float> w(shape, shape+3, strides, data, offset,
   marray::FirstMajorOrder);
\end{verbatim}

The data under a view is accessed similar to the entries of arrays,
i.e.~by coordinates, by single indices, or by means of iterators.
Coordinate permutation works on views exactly the same way it works
on arrays. A sub-view where one coordinates is bound to a certain
value can be obtained as follows:
\begin{verbatim}
marray::View<float> x = w.boundView(2, 1);
// binds dimension 2 to coordinate 1
\end{verbatim}

The member functions \verb$reshape$, \verb$permute$, \verb$transpose$,
\verb$shift$, and \verb$squeeze$ transform the view for which they are
called. They are complemented by member functions called 
\verb$reshapedView$, \verb$permutedView$, etc.~that leave the view for
which they are called unchanged and return a new view that is
transformed in the desired way. The latter functions are first of all
convenient but they also resemble the way transformations are 
implemented in Boost for views whose dimension is fixed at runtime. In 
fact, all operations that change the dimension of a view need to be 
implemented in this way if the dimension of the view is a template 
parameter because the data type changes together with the dimension.

All arithmetic operators are defined on views. Assigning a view
\verb$x$ to a view on mutable data \verb$y$ via \verb$y = x$ copies
the data under \verb$x$ to the memory addressed by \verb$y$, provided
that \verb$x$ and \verb$y$ have the same shape. The copy is performed
per coordinate, not per scalar index or iterator. Potiential memory
overlaps between the two views \verb$x$ and \verb$y$ are taken care of.
Data is copied if necessary, in an assignment \verb$y = x$, as well as 
in in-place operations such as \verb$x += y$.
Assigning a view \verb$x$ to a view on constant data \verb$z$ copies
the view, not the data. This is useful to recycle the memory allocated
for a view on constant data.

In summary, the views, arrays, matrices, and vectors provided in the
Marray package behave exactly like STL containers \cite{austern-1998}
in terms of their fundamental interface. Additional functions going
beyond the interface of STL containers allow the programmer to adjust
the dimension, shape, strides, as well as the storage order at
runtime.
\subsection{Invariants}
For the sake of runtime performance, some redundancy is built into
the view classes. In particular, the size and the shape strides of
views are stored explicitly as attributes although they could be 
computed on demand from the shape and the internal order of 
coordinates. An additional Boolean flag indicates whether a view is 
unstrided and has a zero offset. This flag supports the fast copying 
of data via \verb$memcpy$, provided that views do not overlap. In 
case of overlap, the necessary temporary copy is created internally.

The redundant attributes need to be kept consistent under all possible
transformations of views and arrays. The private member functions 
\verb$testInvariant()$ check for consistency. They are called after 
any transformation in debug mode. The reader is encouraged to look
these functions up in the source code. Since views and arrays are 
fundamental data structures that should work at peak performance in 
released code, it is important that all tests can be removed. A 
function proposed by Stroustrup \cite{stroustrup-2000} is used to meet 
this requirement.
\begin{verbatim}
template<class A> inline void Assert(A assertion) {
   if(!assertion)
      throw std::runtime_error("Assertion failed.");
}
\end{verbatim}
Along with this function, the Boolean constants \verb$NO_DEBUG$ and
\verb$NO_ARG_TEST$ are defined in the namespace \verb$marray$. Invariant
testing and the testing of function arguments is conditioned on these
variables, e.g.
\begin{verbatim}
Assert(NO_DEBUG || this->dimension_ > 0);
Assert(NO_ARG_TEST || std::distance(begin, end) != 0);
\end{verbatim}
In consequence, compilers will remove the respective tests if
\verb$NO_DEBUG$ and \verb$NO_ARG_TEST$ are set to \verb$true$. By
default, both variables are set in accordance with \verb$NDEBUG$.
\subsection{C++0x Extensions}
\label{section:implementation0x}
Features of the C++0x standard proposal \cite{stroustrup-2006}
facilitate three highly desirable extensions whose implementation 
in C++98 would have drawbacks. The C++0x code is part of the Marray
package. However, since C++0x is not yet approved, these extensions
are considered experimental and have to be enabled explicitly by
defining the variables
\begin{verbatim}
HAVE_CPP0X_TEMPLATE_TYPEDEFS
HAVE_CPP0X_VARIADIC_TEMPLATES
HAVE_CPP0X_INITIALIZER_LISTS
\end{verbatim}
\subsubsection{Template Aliases}
Views are declared as class templates in the namespace \verb$marray$:
\begin{verbatim}
template<class T, bool isConst = false> class View;
\end{verbatim}
To support the writing of self-explanatory code, the constants 
\verb$Const = true$ and \verb$Mutable = false$ are defined. Still,
having to write
\begin{verbatim}
marray::View<float, marray::Const> v;
\end{verbatim}
to declare a view on constant data is perhaps not what a programmer
would guess. We could have implemented a class template
\verb$ConstView$ separately. However, even with inheritance, this
would have led to excessive redundancy in the code that would have
made the implementation error prone and hard to maintain \cite{sutter-2004}.
C++0x \cite{stroustrup-2006} provides an elegant solution,
namely the definition of the template alias \cite{reis-2007}
\begin{verbatim}
template<class T> using ConstView = View<T, true>;
\end{verbatim}
This alias allows the programmer to construct a view on constant data
in a straightforward way:
\begin{verbatim}
marray::ConstView<float> v;
\end{verbatim}
\subsubsection{Variadic Templates}
The entries of views and arrays can be accessed by coordinates. For 
the sake of convenience, it should be possible for the programmer to
use \verb$operator()$ with any number of parameters.

C++98 has inherited from C a syntax for functions whose number of
parameters is unspecified at compile time. However, this mechanism is
not type safe \cite{stroustrup-2000} and its use is therefore
discouraged. In the C++98 compatible part of the code, we thus make a 
compromise and implement the operator in a type safe manner for up to
four parameters. A runtime error is issued if the wrong instance is
used. Beyond four dimensions, \verb$operator()$ can be used with one
argument, an iterator to a coordinate sequence.

C++0x defines variadic templates \cite{gregor-2006,gregor-2007} that allow us to
recursively define \verb$operator()$ in a type safe manner for any number of 
parameters. We quote here the main recursive declaration and refer to the 
source code for details.
\begin{verbatim}
template<typename... Args>
   reference_type operator()(const size_t &&,
      const Args && ...);
reference_type operator()(const size_t &&);
\end{verbatim}
\subsubsection{Initializer Lists}
Constructors and member functions of Marray classes take iterators
into coordinate sequences as input. One iterator that points to the
beginning of the sequence is sufficient if the length of the sequence
can be derived, e.g.~in the member function \verb$permute$ of
\verb$View$. Iterator pairs are required otherwise, e.g.~in the member
function \verb$resize$ of \verb$Marray$. Iterators are used excessively
in the STL, so most programmers will find them familiar. However, the use of
iterators and iterator pairs is cumbersome if sequences are known at 
compile time. In fact, neither of the following alternatives is really
convenient:
\begin{verbatim}
size_t shape[] = {4, 2, 3};
marray::Marray<float> a(shape, shape+3);

std::vector<size_t> shape(3);
shape[0] = 4;
shape[1] = 2;
shape[2] = 3;
marray::Marray<float> a(shape.begin(), shape.end());
\end{verbatim}

C++0x defines initializer lists \cite{stroustrup-2007} that allow us 
to overload functions such that the programmer can simply write
\begin{verbatim}
marray::Marray<float> a({4, 2, 3});
\end{verbatim}
\section{Conclusion}
\label{section:conclusion}
We provide C++ class templates for multi-dimensional views and arrays
whose dimension, shape, and size can change at runtime. The C++98
interface of these templates is as convenient as in the best
implementations of arrays with fixed dimensions. Usability is further
improved by C++0x extensions. Our software is free and publicly
available \cite{marray}. We are currently examining different ways of
establishing compatibility with the multi-dimensional arrays that are
native to C and are working on an HDF5 interface.
\section*{Acknowledgements}
The authors thank Daniel Kondermann for fruitful discussions and Oliver
Petra for his contribution to the Marray unit tests.

\end{document}